# Prussian Blue and Prussian Blue Analogues as Emerging Memristive Materials

L. B. Avila[1]

[1]Institute of Condensed Matter and Nanosciences (ICMN), Université catholique de Louvain (UCLouvain), Louvain-la-Neuve B-1348, Belgium.

Abstract.

Prussian blue analogues (PBAs) and related organic materials continue to emerge as promising platforms for next-generation memory and energy-storage technologies due to their rich redox chemistry, ionic mobility, and compatibility with low-cost, scalable fabrication methods. Across the collected studies, electrodeposited Prussian Blue (PB) and Prussian White (PW) thin films consistently exhibit robust resistive switching (RS) with ON/OFF ratios from one to three orders of magnitude, operating in both bipolar and unipolar modes. Structural and spectroscopic analyses confirm that these films display homogeneous composition, well-defined grain boundaries, and ionic pathways that enable filamentary conduction. The switching mechanisms identified through I–V analysis, electrochemical impedance spectroscopy, and quantum transport modeling are governed by ohmic or space-charge-limited conduction, where potassium-ion migration and reversible redox processes drive filament formation and rupture. Furthermore, PB-based devices demonstrate conductance quantization with discrete current steps corresponding to integer and half-integer multiples of $G_0$, indicating ballistic electron transport through atomic-scale channels. Complementary work on perylene-based columnar liquid crystals shows stable, reversible switching enabled by trap-controlled SCLC processes, further broadening the landscape of organic memory materials. Finally, FeHCF/graphene oxide electrodes highlight the versatility of PB analogues in high-voltage symmetric supercapacitors with excellent cycling stability. Collectively, these studies underscore the multifunctionality, scalability, and tunability of PBAs and related systems, positioning them as strong candidates for future ReRAM, neuromorphic computing, multilevel memory, cryptographic hardware, and high-performance energy-storage applications.

Corresponding author email: lindiomar.borges@uclouvain.be

Introduction.

Memristive devices are at the core of several emerging and commercially deployed microelectronic technologies and are expected to play a central role in the future integrated-circuit market [Lanza2022, Ielmini2015, Lanza2025]. Their most immediate impact is in non-volatile memory, where leading semiconductor companies—including TSMC and Intel—are directly incorporating memristive elements into advanced technology nodes for embedded and stand-alone memory solutions [Chiu2019, Chou2018, Chou2020]. In parallel, neuromorphic engineering [Mead1989] is undergoing rapid development, with memristive devices serving as artificial synapses in hardware neural networks designed to accelerate AI training and inference [Aguirre2024, Hui2021, Danilin2015, Mishchenko2022, Perez-Bosch2021, Yu2021, Sebastian2020, Prezioso2015, Zhu2023, Tang2019, Romero-Zaliz2021, Wang2020, Milo2016, Roldan2022]. Beyond memory and neuromorphic computing, memristive devices also play an increasingly important role in cryptographic systems: their inherent stochasticity and device-to-device variability provide high-entropy sources for true random number generators and physical unclonable functions [Wei2016, Gao2022, Lanza2022, Wen2021, Nili2018].

Many memristive systems [Chua1971, Chua1976] exhibit resistive switching (RS), in which the internal conductance can be reversibly modified by several orders of magnitude under electrical stimulus. Resistive Random Access Memories (RRAM), a widely studied subset of memristive devices [Chua2011], rely on ionic transport and localized redox reactions to form and rupture conductive filaments (CFs) within insulating or semiconducting matrices [Spiga2020]. These mechanisms enable both digital (binary) operation for non-volatile memory and analog/multilevel operation, essential for emulating synaptic weight updates in neuromorphic architectures [Tang2019, Roldan2022, Mishchenko2022, Maldonado2023]. Multilevel states can be programmed through voltage-pulse algorithms [Reuben2019, Perez-Bosch2021], compliance-current modulation [Gonzalez-Cordero2019, Poblador2018], or by exploiting quantum transport effects through nanometric constrictions in CFs [Milano2022, Roldan2023a].

Conductance quantization is particularly attractive for multilevel memory and quantum-inspired information processing. Under this regime, device conductance evolves in discrete steps near integer or half-integer multiples of the quantum of conductance, $G_0 = 2e^2/h$. Such behavior emerges when electrons travel ballistically through atomic-scale constrictions—conditions readily achievable in metallic CFs formed during RRAM switching [Miranda2001, Miranda2010]. Since the elastic mean free path of common filament metals (Ag, Cu) is tens of nanometers [Gall2016], nanoconstrictions only a few atoms wide naturally operate far from the diffusive regime. After its first observation in GaAs/AlGaAs heterostructures [VanWees1988], conductance quantization has been reported in atomic contacts [Ohnishi1998], nanotubes [Chico1996], break junctions [He2006], and various dielectric-based RRAM devices [Mehonic2013, Li2015, Milano2022]. Recent studies have highlighted its potential for multilevel memory, neuromorphic computing, and quantum information hardware.

**Prussian Blue and Prussian Blue Analogues as Emerging Memristive Materials**

Prussian Blue (PB) and its analogues (PBAs) constitute a highly promising class of materials for next-generation memristive devices. PB is a mixed-valence iron(III/II) hexacyanoferrate featuring a face-centered cubic structure where $Fe^{3+}$–N≡C–$Fe^{2+}$–C≡N–$Fe^{3+}$ chains extend throughout the lattice. Intervalence charge-transfer transitions between $Fe^{2+}$ and $Fe^{3+}$ sites give rise to PB's characteristic deep-blue coloration and reflect its intrinsic electronic coupling and redox reversibility [Buser1977]. Crucially, the PB lattice contains large interstitial sites capable of accommodating mobile alkali cations (e.g., $K^+$, $Na^+$), which makes PB an inherently ionic-conductive, redox-active material—ideal characteristics for resistive switching.

PB is only one member of a much broader family of Prussian Blue analogues. The general PBA formula,

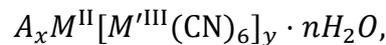

$$A_x M^{II}[M'^{III}(CN)_6]_y \cdot nH_2O,$$

allows extensive chemical tunability: *M* and *M'* can be Fe, Co, Ni, Mn, Cu, Zn, V, or Cr, among others. This modularity enables rational design of PBAs with finely controlled properties including:

- **electronic conductivity and bandgap**
- **ionic mobility and activation energy**
- **redox potentials and electron-transfer pathways**
- **magnetic ordering and spin transitions**
- **electrochromic and structural responses to external stimuli**

A defining feature of PBAs is their **intrinsic vacancy chemistry**: up to one-third of the $[M'(CN)_6]$ octahedra may be absent, creating nanometer-scale voids that enhance ionic diffusion, modify electron transport, and provide flexible redox compensation mechanisms. These vacancies, combined with interstitial cation mobility and hydration dynamics, are central to the fast ionic motion and reversible filament formation observed in PB- and PW-based memristors.

The open-framework architecture of PBAs supports some of the **fastest room-temperature ion mobilities** among solid-state materials. $K^+$ and $Na^+$ ions migrate through the lattice with activation energies in the range of 0.2–0.5 eV, aligning precisely with the requirements for reliable RRAM operation. Additionally, the presence of multiple accessible redox couples (e.g., $Fe^{2+}/Fe^{3+}$, $Co^{2+}/Co^{3+}$, $Mn^{2+}/Mn^{3+}$) allows field-induced electronic bistability, electrochromism, and even metal–insulator transitions under modest bias. Together, these characteristics create ideal conditions for **ionic-drift-driven RS**, multilevel memory operation, and the possibility of coupling electrical, optical, or magnetic degrees of freedom within the same device.

PBAs also offer compelling advantages for **fabrication and integration**. They are composed of earth-abundant, environmentally benign elements and can be synthesized through highly scalable, low-temperature, water-based methods such as electrodeposition, printing, and solution processing. Electrodeposition, in particular, enables nanometer-to-micrometer thickness control, uniform coatings on microelectrodes, and full compatibility with CMOS metal stacks—capabilities difficult to achieve with conventional oxide dielectrics.

Recent studies have demonstrated robust bipolar and unipolar resistive switching in PB, Prussian White (PW), and other PBAs, with conduction dominated by alkali-ion migration and defect-mediated filament formation. Advanced work has revealed **conductance quantization** in PB-based devices, confirming the formation of atomic-scale channels and positioning PBAs as strong candidates for multilevel memory and neuromorphic synaptic elements. Variability analyses further indicate that PB-based memristors possess sufficiently stable switching characteristics for neuromorphic computing and hardware cryptography, including TRNGs and PUFs.

Building upon this rich foundation, the present work focuses on the characterization and modeling of conductance quantization in RRAM devices employing electrodeposited Prussian Blue as the active dielectric layer. Owing to its mixed-valence redox chemistry, open-framework structure, and reversible ionic mobility, PB offers a unique platform for the formation of atomic-scale conduction channels. By analyzing quantized conductance states and elucidating filament dynamics, this study advances the understanding of PB-based memristors and highlights their potential for robust multilevel memory, neuromorphic architectures, and secure hardware applications.

Conclusion.

Taken together, these works demonstrate the broad potential of Prussian Blue, Prussian White, and related hybrid materials as active layers in resistive switching and energy-storage devices. Electrodeposited PB and PW films exhibit stable bipolar and unipolar RS behavior over hundreds of cycles, with conduction dominated by ionic motion primarily potassium ions and defect-mediated filamentary pathways located at grain boundaries. The activation energies, ranging from 0.2 to 0.4 eV, confirm ion-driven transport, while endurance and retention tests show reliable long-term stability suitable for memory applications. Advanced analysis, including EIS and quantum transport modeling, reveals the formation of both nanoscale and atomic-scale conductive filaments, the latter responsible for conductance quantization at multiples of $G_0$, enabling multilevel memory capabilities. Variability studies further indicate that PB-based memristors possess sufficiently low cycle-to-cycle fluctuations for neuromorphic and cryptographic uses such as PUFs and true random number generation. The observed performance across devices combined with low-cost, CMOS-compatible electrodeposition, highlights PBAs as scalable platforms for both nonvolatile memory and secure electronics. Beyond RS devices, PB analogues integrated with reduced graphene oxide achieve high-voltage (2 V), high-stability supercapacitor performance, demonstrating their versatility in electrochemical energy storage. Complementary results from perylene-based liquid crystal systems expand the material space for organic memory devices, revealing stable RS behavior and long-term data retention enabled by reversible molecular reduction. Overall, these findings establish Prussian Blue analogues and related materials as promising, multifunctional candidates for next-generation memory, neuromorphic computing, secure hardware, and high-performance energy-storage technologies, while outlining clear pathways for future optimization in device architecture, environmental stability, and controlled ionic transport.

References.

[1] Avila, L.B.; Müller, C.K.; Hildebrand, D.; Faita, F.L.; Baggio, B.F.; Cid, C.C.P.; Pasa, A.A. *Resistive Switching in Electrodeposited Prussian Blue Layers.* Materials **2020**, *13*, 5618. https://doi.org/10.3390/ma13245618.

[2] Faita, F.L.; Avila, L.B.; Silva, J.P.B.; Boratto, M.H.; Plá Cid, C.C.; Graeff, C.F.O.; Gomes, M.J.M.; Müller, C.K.; Pasa, A.A. *Abnormal Resistive Switching in Electrodeposited Prussian White Thin Films.* Journal of Alloys and Compounds **2022**, *896*, 162971. https://doi.org/10.1016/j.jallcom.2021.162971.

**[3]** Avila, L.B.; Serrano Arambulo, P.C.; Dantas, A.; Cuevas-Arizaca, E.E.; Kumar, D.; Müller, C.K.
*Study on the Electrical Conduction Mechanism of Unipolar Resistive Switching Prussian White Thin Films.*
Nanomaterials **2022**, *12*, 2881. https://doi.org/10.3390/nano12162881.

**[4]** Avila, L.B.; Serrano, P.A.; Quispe, L.T.; Dantas, A.; Costa, D.P.; Arizaca, E.E.C.; Chávez, D.P.P.; Portugal, C.D.V.; Müller, C.K.
*Prussian Blue Anchored on Reduced Graphene Oxide Substrate Achieving High Voltage in Symmetric Supercapacitor.*
Materials **2024**, *17*, 3782. https://doi.org/10.3390/ma17153782.

**[5]** Avila, L.B.; Chulkin, P.; Serrano, P.A.; Dreyer, J.P.; Berteau-Rainville, M.; Orgiu, E.; França, L.D.L.; Zimmermann, L.M.; Bock, H.; Faria, G.C.; Eccher, J.; Bechtold, I.H.
*Perylene-Based Columnar Liquid Crystal: Revealing Resistive Switching for Nonvolatile Memory Devices.*
Journal of Molecular Liquids **2024**, *402*, 124757. https://doi.org/10.1016/j.molliq.2024.124757.

**[6]** Avila, L.B.; Cantudo, A.; Villena, M.A.; Maldonado, D.; Abreu Araujo, F.; Müller, C.K.; Roldán, J.B.
*Variability Analysis in Memristors Based on Electrodeposited Prussian Blue.*
Microelectronic Engineering **2025**, *300*, 112376. https://doi.org/10.1016/j.mee.2025.112376.

**[7]** Cantudo, A.; Avila, L.B.; Villena, M.A.; Jiménez-Molinos, F.; Ducarme, C.; Lopes Temporao, A.; Moureaux, A.; Abreu Araujo, F.; Müller, C.K.; Roldán, J.B.
*Conductance Quantization in Memristive Devices with Electrodeposited Prussian Blue-Based Dielectrics.*
Materials Science in Semiconductor Processing **2026**, *203*, 110253. https://doi.org/10.1016/j.mssp.2025.110253